\documentclass[10pt, conference, compsocconf]{IEEEtran}

\usepackage{cite}
\ifCLASSINFOpdf
  \usepackage[pdftex]{graphicx}
          \else
        \usepackage[dvips]{graphicx}
          \fi
\usepackage[cmex10]{amsmath}
\usepackage{algpseudocode}
\usepackage{array}
\usepackage{mdwmath}
\usepackage{mdwtab}
\usepackage{eqparbox}
\usepackage[tight,footnotesize]{subfigure}
\usepackage{fixltx2e}
\usepackage{stfloats}
\usepackage{url}
\usepackage{booktabs}
\usepackage{amssymb}
\newcommand{\mypar}[1]{\textbf{#1: }}
\usepackage{forloop}
\newcounter{ct}
\newcommand{\markdent}[1]{\forloop{ct}{0}{\value{ct} < #1}{\hspace{\algorithmicindent}}}
\newcommand{\markcomment}[1]{\Statex\markdent{#1}}
\newcommand{\pushright}[1]{\ifmeasuring@#1\else\omit\hfill$\displaystyle#1$\fi\ignorespaces}
\newcommand{\pushleft}[1]{\ifmeasuring@#1\else\omit$\displaystyle#1$\hfill\fi\ignorespaces}
\makeatother

\newcolumntype{K}[1]{>{\centering\arraybackslash}p{#1}}
\usepackage[colorinlistoftodos,prependcaption,textsize=small]{todonotes}

\usepackage{multicol}
\usepackage{dsfont}

\usepackage{ mathrsfs }

\usepackage{float}
\newfloat{code}{ht}{lop}
\floatname{code}{Code}

\newcommand{\step}[1]{\mathds{1}_{#1}}
\newcommand{\algsmall}{\small}
\newcommand{\mathsmall}{\small}
\newcommand{\tablesize}{\small}
\newcommand{\pida}{\Pi_\mathrm{2D}}
\newcommand{\pidb}{\Pi_\mathrm{3D}}
\newcommand{\pidc}{\Pi_\mathrm{4D}}
\def\bsq#1{\lq{#1}\rq}

\begin{document}
\title{Communication-Avoiding Parallel Algorithms for\\ Solving Triangular Systems of Linear Equations}
\author{
    \IEEEauthorblockN{Tobias Wicky}
    \IEEEauthorblockA  {
        Department of Computer Science\\
        ETH Zurich\\
        Zurich, Switzerland\\
        \texttt{twicki{@}ethz.ch}
    }
    \and
    \IEEEauthorblockN{Edgar Solomonik}
    \IEEEauthorblockA  {
        Department of Computer Science\\
        University of Illinois at Urbana-Champaign\\
        Urbana, IL, USA\\
        \texttt{solomon2@illinois.edu}
    }
    \and
    \IEEEauthorblockN{Torsten Hoefler}
    \IEEEauthorblockA  {
        Department of Computer Science\\
        ETH Zurich\\
        Zurich, Switzerland\\
        \texttt{htor@inf.ethz.ch}
    }
}

\maketitle

\begin{abstract}
We present a new parallel algorithm for solving triangular systems with multiple right hand sides (TRSM).
TRSM is used extensively in numerical linear algebra computations, both to solve triangular linear systems of equations as well as to compute factorizations with triangular matrices, such as Cholesky, LU, and QR.
Our algorithm achieves better theoretical scalability than known alternatives, while maintaining numerical stability, via selective use of triangular matrix inversion.
We leverage the fact that triangular inversion and matrix multiplication are more parallelizable than the standard TRSM algorithm.
By only inverting triangular blocks along the diagonal of the initial matrix, we generalize the usual way of TRSM computation and the full matrix inversion approach.
This flexibility leads to an efficient algorithm for any ratio of the number of right hand sides to the triangular matrix dimension.
We provide a detailed communication cost analysis for our algorithm as well as for the recursive triangular matrix inversion.
This cost analysis makes it possible to determine optimal block sizes and processor grids a priori.
Relative to the best known algorithms for TRSM, our approach can require asymptotically fewer messages, while performing optimal amounts of computation and communication in terms of words sent.
\end{abstract}


\IEEEpeerreviewmaketitle

\section{Introduction}
\label{sec:intro}
Triangular solve for multiple right hand sides (TRSM) is a crucial subroutine in many numerical linear algebra algorithms, such as LU and Cholesky factorizations~\cite{gustavson1997recursion,solomonik2011communication}.
Moreover, it is used to solve linear systems of equations once the equation matrix is decomposed using any factorization involving a triangular matrix factor. 
We consider TRSM for dense linear equations in the matrix form,
{\mathsmall
\[L \cdot X = B,\]
}\noindent
where $\left.L \in \mathbb{R}^{n \times n}\right. $ is a lower-triangular matrix while $\left.B \in \mathbb{R}^{n \times k}\right. $, $\left.X \in \mathbb{R}^{n \times k}\right. $ are dense matrices.
We study the communication cost complexity of two variants of the TRSM algorithm for parallel execution on $p$ processors with unbounded memory.
First, we present an adaptation of a known recursive scheme for TRSM~\cite{gustavson} along with a complete communication cost analysis.
Then, we demonstrate a new algorithm that uses selective triangular inversion to reduce the synchronization cost over known schemes, while preserving optimal communication and computation costs.
Careful choice of algorithmic parameters, allows us to achieve better asymptotic complexity for a large (and most important) range of input configurations.

Our TRSM algorithm leverages matrix multiplication and triangular matrix inversion as primitives.
We provide a communication-efficient parallel matrix multiplication algorithm that starts from a 2D cyclic distribution, a modest enhancement to existing approaches.
Triangular matrix inversion provides the key ingredient to the lower synchronization cost in our TRSM algorithm.
Unlike general matrix inversion, triangular inversion is numerically stable~\cite{du1992stability} and can be done with relatively few synchronizations.
We present a known parallel approach for triangular inversion and provide the first communication cost analysis thereof.

We invert triangular diagonal blocks of the $L$ matrix at the start of our TRSM algorithm, increasing the computational granularity of the main part of the solver.
Inverting blocks, rather than the whole matrix, also allows our algorithm to be work-efficient in cases when the number of right-hand sides is smaller than the matrix dimension.
We formulate the algorithm in an iterative, rather than a recursive manner, avoiding overheads incurred by the known parallel recursive TRSM approach.
This innovation reduces communication cost by a factor of $\Theta(\log(p))$ relative to the recursive algorithm when the number of right-hand sides is relatively small.
At the same time, across a large range of input parameters, we achieve a synchronization cost improvement over the recursive approach by a factor of $\Theta\left(\left(\frac{n}{k}\right)^{1/6}p^{2/3}\right) $.

\section{Preliminaries}
\label{sec:prelims}
\subsection{Execution Time Model}
The model we use to calculate the parallel execution time of an algorithm along its critical path is the $\alpha-\beta-\gamma$ model.
It describes the total execution time of the algorithm $T$ in terms of the floating point operations (flop) count $F$, the bandwidth $W$ (number of words of data sent and received) and the latency $S$ (number of messages sent and received) along the critical path~\cite{SCKD_TECHREP_2014} in the following fashion:
{\mathsmall
\[T=  \alpha \cdot S+ \beta \cdot W +\gamma\cdot F.\]}\noindent
For all the terms we only show the leading order cost in terms of $n,k$ and $p$.
We assume that every processor can send and receive one message at a time in point to point communication.
We do not place constraints on the local memory size.

\subsection{Notation}
For brevity, we will sometimes omit specification of the logarithm base, using $\log$ to denote $\log_2$.
We will make frequent use of the unit step function
{\mathsmall 
 \[\step{x} = \begin{cases} 1 : x > 1 \\ 0 : x \leq 1. \end{cases}\] 
}\noindent
We use $\Pi(x_1,\ldots,x_n)$to denote single processors in an n-dimensional processor grids. 

To refer to successive elements or blocks, we will use the colon notation where 
 \( i:j = [i, i+1, \dots, j-1] \quad i < j.\)
To refer to strided sets of elements or blocks, we will write 
{\mathsmall 
 \[i:k:j = [i, i+k, i+2k, \dots, i + \alpha k] \quad \max \alpha\text{ s.t. } \alpha k < j.\] 
}\noindent
The colon notation can be applied to a list and should there be considered element-wise.
To use subsets of processors, we use the $\circ$ notation in the way that
 \(\Pi(x,\circ,z) = \Pi(x,1:p_\mathrm{y},z)\)
denotes a (1-dimensional) grid of processors in the y-dimension.

Global matrices are denoted by capital letters whereas locally owned parts have square brackets:
If $L$ is distributed on $\Pi(\circ, \circ,1)$, every processor $\Pi(x,y,1)$ owns $L[x,y]$.
Matrix elements are accessed with brackets.

\subsection{Previous Work}
We now cover necessary existing building blocks (e.g. collective communication routines) and previous work.
In particular, we overview related results on communication cost of matrix multiplication and triangular solves.

\subsubsection{Collective Communication} 
\label{sec:coll}
In~\cite{geijn}, Chan et al. present a way to perform reduction, allreduction and broadcast via allgather, scatter, gather and reduce-scatter.
The latter set of collectives can be done using recursive doubling (butterfly algorithms)~\cite{geijn,rec_double_and_half} for a power of two number of processors.
If we have a non-power of two number of processors, the algorithm described in~\cite{bruck_algo} can be used.
For simplicity, we do not consider reduction and broadcast algorithms that can achieve a factor of two less in cost in specific regimes of $\alpha$ and $\beta$~\cite{traff2008optimal}.
If we use butterfly methods, the cost of an allgather of $n$ words among $p$ processors is
{\mathsmall 
 \[T_\mathrm{allgather}(n,p)=\alpha \cdot \log p + \beta \cdot n\step{p}.\] 
}\noindent
Scatter and gather also have the same cost~\cite{rec_double_and_half}, 
{\mathsmall
\begin{align*}
T_\mathrm{scatter}(n,p)&=\alpha \cdot \log p + \beta \cdot n\step{p},\\
T_\mathrm{gather}(n,p)&=\alpha \cdot \log p + \beta \cdot n\step{p}. 
\end{align*}
}\noindent
Reduce-scatter uses the same communication-path and has same communication cost but we have to add the additional overhead of local computation,
{\mathsmall 
 \[T_\mathrm{reduce-scatter}(n,p)=\alpha \cdot \log p + \beta \cdot n\step{p} + \gamma \cdot n\step{p}.\] 
}\noindent
The cost of an all-to-all among $p$ processors is 
{\mathsmall 
 \[T_\mathrm{alltoall}(n,p) = \alpha\cdot \log(p) + \beta\cdot \frac{n\log p}{2}.\] 
}\noindent
The combination of the algorithms leads to the following costs for reduction, allreduction, and broadcast:
{\mathsmall
\begin{align*}
T_\mathrm{reduction}(n,p)&=\alpha \cdot 2\log p + \beta \cdot 2n\step{p} + \gamma \cdot n \step{p},\\
T_\mathrm{allreduction}(n,p)&=\alpha \cdot 2\log p + \beta \cdot 2n\step{p} + \gamma \cdot n \step{p},\\
T_\mathrm{bcast}(n,p)&=\alpha \cdot 2\log p + \beta \cdot 2n\step{p}.
\end{align*}
}\noindent
\subsubsection{Matrix Multiplication}
Communication-efficient parallel algorithms for matrix multiplication have been analyzed extensively~\cite{dekel:657,matmul3d,snirmatmul,berntsen1989communication,mccol_tiskin_99,Johnsson:1993:MCT:176639.176642}.
In~\cite{carma}, Demmel et al. present algorithms that are asymptotically optimal for matrix multiplication of arbitrary (potentially non square) matrices.
If we neglect the memory terms, their work shows that matrix multiplication can be done with the following costs.

\mypar{Bandwidth} When multiplying a matrix $A$ that is of dimension $\left.n \times n\right. $ with a matrix $B$ of dimensions $\left.n \times k\right. $ with $p$ processors, we obtain an asymptotic bandwidth cost of 
{\mathsmall
\begin{equation*}
{\small
W_\mathrm{MM}(n,k,p) =
\left\{ \begin{alignedat}{4}
  &\mathcal{O}\left(\frac{nk}{\sqrt{p}}\right)   &n&>k\cdot \sqrt{p} \\
  &\mathcal{O}\left(\left(\frac{n^2k}{p}\right)^{2/3}\right) \hspace*{5mm}k/p\leq &n& \leq k\cdot \sqrt{p}\\
  &\mathcal{O}\left(n^2\right)  &n&<k/p.
        \end{alignedat}\right. 
}
\end{equation*}
}\noindent
We refer to the first of the cases of $W_\mathrm{MM}$, as the case of two large dimensions, here the matrix $A$ is much larger than the right hand side $B$, when the best way of performing a matrix multiplication is to use a two dimensional layout for the processor grid.
The second case, three large dimensions, has matrices $A$ and $B$ of approximately the same size.
A three dimensional grid layout is optimal here.
And the third case, one large dimension, is the case where the right hand side $B$ is larger than the triangular matrix $A$, the best way to do a matrix multiplication is to use a one dimensional layout for the processor grid.

\mypar{Latency} Assuming unlimited memory, matrix multiplication as presented in~\cite{carma} can be done with a latency of 
{\mathsmall
$$S_\mathrm{MM}(p) = \mathcal{O}\left(\log(p)\right).$$
}\noindent
\mypar{Flop Cost} Matrix multiplication takes $\mathcal{O}\left(n^2k\right)$ flops, which can be divided on $p$ processors and therefore we have
{\mathsmall 
 \[F_\mathrm{MM}(n,k,p) = \mathcal{O}\left(\frac{n^2k}{p}\right).\] 
}\noindent
\mypar{Previous Analysis} For the case where $\left.k=n\right.$ the bandwidth analysis of a general matrix multiplication goes back to what is presented in~\cite{snirmatmul}.
Aggarwal et al. present a cost analysis in the LPRAM model. 
In that work, the authors show that the same cost can also be achieved for the transitive closure problem that can be extended to the problem of doing an LU decomposition.
The fact that these bandwidth costs can be obtained for the LU decomposition was later demonstrated by Tiskin~\cite{tiskin2002bulk}.
He used the bulk synchronous parallel (BSP) execution time model.
Since the dependencies in LU are more complicated than they are for TRSM, we also expect TRSM to be able to have the same asymptotic bandwidth and flop costs as a general matrix multiplication.

\subsubsection{Triangular Matrix Solve for Single Right Hand Sides}
Algorithms for the problem of triangular solve for a single right hand side (when $X$ and $B$ are vectors) have been well-studied.
A communication-efficient parallel algorithm was given by Heath and Romine~\cite{heath1988parallel}.
This parallel algorithm was later shown to be an optimal schedule in latency and bandwidth costs via lower bounds~\cite{SCKD_TECHREP_2014}.
However, when $X$ and $B$ are matrices $\left(k>1\right) $, it is possible to achieve significantly lower communication costs relative to the amount of computation required.
The application of selective inversion has been used to accelerate repeated triangular solves that arise in preconditioned sparse iterative methods~\cite{raghavan1998efficient}.

\subsubsection{Recursive Triangular Matrix Solve for Multiple Right Hand Sides}
\label{sec:prevtrsm}
A recursive approach of solving the TRSM-problem was presented in the work of Elmroth et al.~\cite{gustavson}.
The initial problem, \(\left.L \cdot X = B\right. \) can be split into
\(L \cdot \begin{bmatrix}
           X_1 & X_2
          \end{bmatrix}
          = \begin{bmatrix}
             B_1 & B_2
            \end{bmatrix},
\)
which yields two independent subproblems:
{\mathsmall
\begin{align*}
 L \cdot X_1 = B_1, \quad \quad
 L \cdot X_2 = B_2.
\end{align*}
}\noindent
hence the subproblems are independent and can be solved in parallel.

The other, dependent splitting proposed divides the triangular matrix, yielding two subtasks that have to be solved one at a time
{\mathsmall
\[\begin{bmatrix}
   L_{11} \\
   L_{12} & L_{22}
  \end{bmatrix} \cdot \begin{bmatrix}
  X_1\\
  X_2
  \end{bmatrix} = \begin{bmatrix}
    B_1\\
    B_2
    \end{bmatrix},
\]
}\noindent
where we obtain the dependent subproblems:
{\mathsmall
\begin{align*}
 L_{11} \cdot X_1 &= B_1 \quad \text{and} \quad  L_{22} \cdot X_2 = B_2 - L_{12}\cdot X_1.
\end{align*}}\noindent
These problems are dependent as we need the solution $X_1$ to solve the second problem.

Parallel TRSM algorithms with 3D processor grids can reduce the communication cost in an analogous fashion to matrix multiplication.
Irony and Toledo~\cite{irony_3d_TRSM} presented the first parallelization of the recursive TRSM algorithm with a 3D processor grid.
They demonstrated that the communication volume of their parallelization is $\mathcal{O}(nkp^{1/3}+n^2p^{1/3})$.
Thus each processor communicates $\mathcal{O}((nk+n^2)/p^{2/3})$ elements, which is asymptotically equal to $W_\mathrm{MM}(n,k,p)$ when $k=\Theta(n)$.
However, they did not provide a bound on the latency cost nor on the communication bandwidth cost along the critical path,
so it is unclear to what extent the communication volume is load-balanced.
Lipshitz~\cite{lipshitz_2013} provides an analysis of the recursive TRSM algorithm in the same communication cost model as used in this paper.
For the case of $k=n$, his analysis demonstrates
{\mathsmall 
 \[T_\mathrm{TRSM-L}(n,n,p)=\mathcal{O}(p^{2/3}\cdot\alpha+ n^2/p^{2/3}\cdot \beta + n^3/p\cdot \gamma).\] 
}\noindent
For some choices of algorithmic parameters, the analysis in~\cite{lipshitz_2013} should lead to the bandwidth cost,
{\mathsmall 
 \[W_\mathrm{TRSM-L}(n,k,p)=\mathcal{O}\bigg(\Big(\frac{n^2k}{p}\Big)^{2/3}\bigg),\] 
}\noindent
which is as good as matrix multiplication, $W_\mathrm{MM}(n,k,p)$.
However, it is unclear how to choose the parameters of the formulation in~\cite{lipshitz_2013} to minimize latency cost for general $n,k$.
Prior to presenting our main contribution (an inversion-based TRSM algorithm with a lower latency cost), we provide a simpler form of the recursive TRSM algorithm and its asymptotic cost complexity.
Inversion has previously been used in TRSM implementations~\cite{tomov2010dense}, but our study is the first to consider communication-optimality.
We start by presenting a subroutine for 3D matrix multiplication which operates from a starting 2D distribution, simplifying the subsequent presentation of TRSM algorithms.

\section{Matrix Multiplication}
\label{sec:new_trmm}
{\algsmall
\rule{\columnwidth}{.5pt}
\vspace{-.2in}

\(B = \text{\bf MM}(L,X,\pida,n,k,p,p_1,p_2)\)

\vspace{-.15in}
\rule{\columnwidth}{.5pt}
\begin{algorithmic}[1]
\Require
\Statex The processor grid $\pida$ has dimensions $\sqrt{p}\times \sqrt{p}$ 
\Statex $L$ is an $n\times n$ matrix, distributed on $\pida$ in a cyclic layout, so $\pida(x,y)$ owns $L[x,y]$ of size $\left.\frac{n}{\sqrt{p}} \times \frac{n}{\sqrt{p}}\right. $ such that 
\Statex $L[x,y](i,j)=L(i\sqrt{p}+x,j\sqrt{p}+y)$.
\Statex $X$ is a dense $n \times k$ matrix is distributed cyclically so that $\pida(x,y)$ owns $X[x,y]$ of size $\frac{n}{\sqrt{p}}\times \frac{k}{\sqrt{p}}$ 
\Statex
\State Define a $p_1\times \sqrt{p_2}\times p_1\times \sqrt{p_2}$ processor grid $\pidc$, such that $\pidc(x_1,x_2,y_1,y_2)=\pida(x_1+p_1x_2,y_1+p_2y_2)$ owns blocks $L[x_1,x_2,y_1,y_2]$ and $X[x_1,x_2,y_1,y_2]$
\State $L'[x_1,y_1] = \textbf{Allgather}\left(L[x_1,\circ,y_1,\circ],\pidc(x_1,\circ,y_1,\circ)\right)$ \label{li:allgL}
\State $X'[x_1,y_1,x_2,y_2]$ \label{li:tX1}
\Statex $\quad=\textbf{Transpose}(X[x_1,x_2,y_1,y_2],\pidc(x_1,x_2,y_1,y_2),x_2,y_1)$
\State $X''[y_1,x_1,x_2,y_2]$ \label{li:tX2}
\Statex $\quad=\textbf{Transpose}(X'[x_1,y_1,x_2,y_2],\pidc(x_1,x_2,y_1,y_2),x_1,y_1)$
\State $X'''[y_1,x_2,y_2] = \label{li:allgX}$
\markcomment{1}{$\textbf{Allgather}(X''[y_1,\circ,x_2,y_2],\pidc(\circ,x_2,y_1,y_2))$}
\State $\pidc(x_1,x_2,y_1,y_2)$ :  \label{li:mm}
\Statex $\quad B''[x_1,y_1,x_2,y_2]=L'[x_1,y_1]\cdot X'''[y_1,x_2,y_2]$
\State $B'[x_1,y_1,x_2,y_2]$ \label{li:srB}
\Statex $\quad=\textbf{Scatter-reduce}(B''[x_1,\circ,x_2,y_2],\pidc(x_1,x_2,\circ,y_2))$
\State $B[x_1,x_2,y_1,y_2]$ \label{li:tB}
\Statex $\quad=\textbf{Transpose}(B'[x_1,y_1,x_2,y_2],\pidc(x_1,x_2,y_1,y_2),x_2,y_1)$
\Ensure
\Statex $B=LX$ is distributed the same way as $X$
\end{algorithmic}
\vspace{-.2in}
\rule{\columnwidth}{.5pt}
}
We present an algorithm for 3D matrix multiplication~\cite{dekel:657,matmul3d,snirmatmul,berntsen1989communication,mccol_tiskin_99,Johnsson:1993:MCT:176639.176642} that works efficiently given input matrices distributed cyclically on a 2D processor grid.
The algorithm is well-suited for the purposes of analyzing TRSM algorithms.
We define the algorithm using a $p_1\times \sqrt{p_2}\times p_1\times \sqrt{p_2}$ processor grid, where $\sqrt{p}=p_1\sqrt{p_2}$ in order to provide well-defined transitions from a distribution on a $\sqrt{p}\times \sqrt{p}$ processor grid to faces of a 3D $p_1\times p_1\times p_2$ processor grid.
The latter 3D processor grid is being used implicitly in our construction.
The algorithm assumes divisibility among $p,p_1,p_2$ and $\sqrt{p_2}$.

\subsection{Cost Analysis of the 3D Matrix Multiplication Algorithm}

We analyze the algorithm with account for constant factors in the key leading order costs.
The communication costs incurred at line~\ref{li:allgL},~\ref{li:allgX}, and~\ref{li:srB} correspond to the cost of respective collectives, given in Section~\ref{sec:coll}.

The transpose on line~\ref{li:tX2} always occurs on a square processor grid, and so involves only one send and receive of a block.
The transposes on line~\ref{li:tX1} and line~\ref{li:tB} are transposes on 2D grids of $p_1\times \sqrt{p_2}$ processors, with each processor owning $nk/p$ elements.
The cost of these transposes is no greater than an all-to-all among $\sqrt{p}$ processors, which can be done with cost $\mathcal{O}(\alpha\cdot \log(p) + \beta\cdot nk\log(p)/p)$.
We consider only the asymptotic cost of this transpose, since it will be low order so long as $p_1\gg 1$.
Based on the above arguments, the cost for MM is given line-by-line in the following table.

{\tablesize
\centering
\begin{tabular}{c | l }
Line~\ref{li:allgL} 
& $\alpha\cdot \log(p_2)+\beta\cdot \frac{n^2}{p_1^2}\step{p_2}$ 
\vspace{.01in} \\ \hline
Line~\ref{li:tX1} & $\mathcal{O}\Big(\alpha\cdot \log(p)+\beta\cdot \frac{nk\log(p)}{p}\Big)$ \\ \hline
Line~\ref{li:allgX} & $\alpha\cdot \log(p_1)+\beta\cdot \frac{nk}{p_1p_2}$ 
\vspace{.01in} \\ \hline
Line~\ref{li:tX2} & $\alpha+\beta\cdot \frac{nk}{p}$ 
\vspace{.01in}\\ \hline
Line~\ref{li:mm} & $\gamma\cdot \frac{n^2k}{p}$
\vspace{.01in}\\ \hline
Line~\ref{li:srB} & $\alpha\cdot \log(p_1) + (\beta+\gamma)\cdot \frac{nk}{p_1p_2}$
\vspace{.01in} \\ \hline
Line~\ref{li:tB} & $\alpha\cdot l + \beta\cdot \frac{nkl}{p}$ 
\end{tabular}

}\noindent
To leading order, the cost of MM is given by
{\mathsmall
\begin{align*}
T_\mathrm{MM}(n,k,p,p_1,p_2)=& \beta\cdot \Big(\frac{n^2}{p_1^2}\step{p_2}+\frac{2nk}{p_1p_2}\Big)+\gamma\cdot \frac{n^2k}{p} \\
&+\mathcal{O}\bigg(\alpha\cdot \log(p)+\beta\cdot \frac{nk\log(p)}{p}\bigg).
\end{align*}
}\noindent
The last communication cost term (due to the rectangular grid transpose) is only of leading order when $p_1\approx \log(p)$.
A square processor grid is not a good initial/final layout for $X$ and $B$ in this case, and the problem would be addressed by choosing an alternative one.
We disregard this issue, because we will use the algorithm only for $n\geq k$.

\section{Recursive TRSM}
\label{sec:oldtrsm}
We provide a recursive TRSM algorithm for solving $LX=B$ using the techniques covered in Section~\ref{sec:prevtrsm}.
Our algorithm works recursively on a $p_r\times p_c$ processor grid.
We will define the processor grid to be square $\left(p_r=p_c\right)$ when $n\geq k$, but rectangular $\left(p_r<p_c\right) $ when $n<k$.
So long as $p<k/n$, we will choose $p_c=(k/n)p_r$.
This strategy implies the largest of the matrices $L$ and $B$ will be partitioned initially so each cyclically-selected block is close to square.
The algorithm starts by partitioning the processor grid into $p_c/p_r$ square grids, if $p_c>p_r$, replicating the matrix $L$ and computing a subset of $kp_r/p_c$ columns of $X$ on each.
Then the algorithm partitions $L$ into $n/2\times n/2$ blocks recursively, executing subproblems with all processors.
At a given threshold, $n_0$, the algorithm stops recursing, gathers $L$ onto all processors, and computes a subset of columns of $X$ with each processor.

{\algsmall
\rule{\columnwidth}{.5pt}
\vspace{-.2in}

\(X=\text{Rec-TRSM}(L,B,\pida,n,k,p_r,p_c,n_0)\)

\vspace{-.15in}
\rule{\columnwidth}{.5pt}
\begin{algorithmic}[1]
 \Require
\Statex $L$ is a lower triangular $n\times n$ matrix, distributed on $p_r\times p_c$ in a cyclic layout, so $\pida(x,y)$ owns $L[x,y]$ of size $\frac{n}{p_r}\times \frac{n}{p_c}$ such that $L[x,y](i,j)=L(ip_r+x,jp_c+y)$.
\Statex $B$ is a dense $n \times k$ matrix is distributed cyclically so that $\pida(x,y)$ owns $X[x,y]$ of size $\frac{n}{p_r}\times \frac{k}{p_c}$ 
  \If{$p_r = qp_c$ and $q>1$}
    \State Define $p_r\times p_r\times q$ processor grid $\pidb$, such that 
    \markcomment{1}{$\pidb(x,y,z)=\pida(x_1,y+p_rz)$ owns blocks}
    \markcomment{1}{$L[x,y,z]$, and $B[x,y,z]$}
    \State $L'[x,y] = \textbf{Allgather}(L[x,y,\circ],\pidb(x,y,\circ))$ \label{li2:allgL}
    \State $X[\circ,\circ,z] = \textbf{Rec-TRSM}(L'[\circ,\circ],B[\circ,\circ,z],$
    \markcomment{5}{$\pida(\circ,\circ,z),n,k/q,p_r,p_r,n_0)$}
  \ElsIf {$n\leq n_0$ or $p_r=p_c=1$} 
    \State $L = \textbf{Allgather}(L[\circ,\circ],\pida(\circ,\circ))$
    \State $B[x+yp_r] = \textbf{AllToAll}(B[\circ,y],\pida(\circ,y))$
    \State $\pida(x,y)$ : $X[x+yp_r]=L^{-1}B[x+yp_r]$ 
    \State $X[x,y] = \textbf{AllToAll}(B[x+\circ p_r],\pida(\circ,y))$
  \Else
    \State Partition 
           $L=\begin{bmatrix}L_{11} & 0 \\ L_{21} & L_{22} \end{bmatrix}$ so $L_{ij}\in\mathbb{R}^{\frac n2 \times \frac n2}$
    \State Partition 
           $B=\begin{bmatrix}B_{1} \\ B_{2} \end{bmatrix}, 
           X=\begin{bmatrix}X_{1} \\ X_{2} \end{bmatrix}$, so
          $B_{i},X_i\in\mathbb{R}^{\frac n2 \times k}$
    \State $X_1=\textbf{Rec-TRSM}(L_{11},B_1,\pida,n/2,k,p,p_r,p_c,n_0)$.
    \State $B'_2=B_2-\textbf{MM}(L_{21},X_1,\pida,n/2,k,p,$ 
    \Statex $\quad\quad\quad\quad\quad\quad\quad\quad\quad p^{1/3}(n/k)^{1/3},p^{1/3}(n/k)^{2/3})$.
    \State $X_2=\textbf{Rec-TRSM}(L_{22},B'_2,\pida,n/2,k,p,p_r,p_c,n_0)$. 
  \EndIf
 \Ensure
    \Statex $X=L^{-1}B$ is distributed on $\pida$ in the same way as $B$
\end{algorithmic}
\vspace{-.2in}
\rule{\columnwidth}{.5pt}
}
\subsection{Cost Analysis of the Recursive Algorithm}
\label{sec:oldcost}

We select $p_c=\max(\sqrt{p},\min(p,\sqrt{pk/n}))$ and $p_r=p/p_c=\min(\sqrt{p},\max(1,\sqrt{pn/k}))$.
The cost of the allgather on line~\ref{li2:allgL} is
{\mathsmall 
\[T_\mathrm{part-cols}(n,p_r)=\mathcal{O}\bigg(\beta\cdot \frac{n^2}{p_r^2} + \alpha\cdot \log(p)\bigg),\] 
}\noindent
since each $L'[x,y]\in\mathbb{R}^{n/p_r\times n/p_r}$ and is lower triangular.
Once we have a square processor grid, we partition $L$, yielding the recurrence,
{\mathsmall 
\begin{align*}
T_\mathrm{RT}(n,k,p,n_0)=&T_\mathrm{MM}\Big(n/2,k,p,p^{1/3}\big(\frac nk\big)^{1/3},p^{1/3}\big(\frac nk\big)^{2/3}\Big) \\
&+ 2T_\mathrm{RT}(n/2,k,p,n_0).
\end{align*}
}\noindent
We now derive the cost of the algorithm for different relations between $n$, $k$, and $p$, as in the expression for $T_\mathrm{MM}$.

\textbf{One large dimension: } 
When $n<k/p$, we have $p_r=1$ and $p_c=p$ and the first allgather will be the only communication,
therefore,
{\mathsmall 
\begin{equation*}
T_\mathrm{RT1D}(n,k,p)=\mathcal{O}\bigg(\alpha\cdot \log(p) + \beta\cdot n^2  + \gamma\cdot \frac{n^2k}{p}\bigg).
\end{equation*}
}\noindent
\textbf{Two large dimensions: } 
When $n>k\sqrt{p}$, we will select $p_r=p_c=\sqrt{p}$ and the column partitioning of $B$ is not performed.
In this case, the MM algorithm will always have $p_2=1$ (it will be 2D).
For sufficiently large $k$, this leads us to the recurrence,
{\mathsmall
\begin{align*}
T_\mathrm{RT2D}(n,k,p,n_0)=&T_\mathrm{MM2D}(n/2,k,p) + 2T_\mathrm{RT2D}(n/2,k,p,n_0),
\end{align*}
}\noindent
where 
{\mathsmall
\(T_\mathrm{MM2D}(n,k,p)= \mathcal{O}\left(\alpha\cdot \log(p)+\beta\cdot \frac{nk}{\sqrt{p}}+\gamma\cdot \frac{n^2k}{p} 
\right).\)
}\noindent
The bandwidth cost stays the same at every recursive level, while the computation cost decreases by a factor of $2$.
At the base case, we incur the cost $T_\mathrm{RTBC}(n_0,k,p)=$
{\mathsmall 
\[\mathcal{O}\bigg(\alpha\cdot \log(p)+\beta\cdot \Big(n_0^2+\frac{n_0k\log(p)}{p}\Big) +\gamma\cdot \frac{n_0^2k}{p} \bigg)\] 
}\noindent
We select $n_0=\max(\sqrt{p},n\log(p)/\sqrt{p})$, so $n/n_0\leq \sqrt{p}/\log(p)$, which results in the overall cost,
{\mathsmall 
\begin{align*}
T_\mathrm{RT2D}(n,k,p)=\mathcal{O}\bigg( \alpha\cdot \sqrt{p}+\beta\cdot \frac{nk\log(p)}{\sqrt{p}}+\gamma\cdot\frac{n^2k}{p}\bigg).
\end{align*}
}\noindent
The bandwidth cost above is suboptimal by a factor of $\mathcal{O}(\log(p))$.
The overhead is due to the recursive algorithm re-broadcasting some of the same elements of $L$ at every recursive level.
We use an iterative approach for our subsequent TRSM algorithm to avoid this redundant communication.

\textbf{Three large dimensions: } 

When $\left.k/p<n<k/\sqrt{p}\right. $, the algorithm partitions the columns of $B$ initially then recursively partitions $L$.
In particular, we select $p_c = \max(\sqrt{p},\sqrt{pk/n})$ and $p_r=p/p_c=\min(\sqrt{p},\sqrt{pn/k})$, so the first step partitions a rectangular processor grid into $\max(1,k/n)$ fewer processor grids.
After the first step, which partitions $B$, we have independent subproblems with $p_r$ processors.
We now start recursively partitioning $L$, yielding the cost recurrence,
{\mathsmall
\begin{align*}
T&_\mathrm{RT3D}(n,k,p_r^2,n_0)= T_\mathrm{MM3D}(n/2,k,p_r^2) \\
&+ T_\mathrm{part-cols}(n,p)\step{k/n}+2T_\mathrm{RT3D}(n/2,k,p_r^2,n_0).
\end{align*}
}\noindent
Above, we always employ the MM algorithm in the 3D regime by selecting
$p_1=p_r^{2/3}(n/l)^{1/3}$ and $p_2=p_r^{2/3}(n/l)^{2/3}$ where $l=kp_r/p_c$.
In this case, $T_\mathrm{MM}$ reduces to $T_\mathrm{MM3D}(n,k,p)=$
{\mathsmall
\begin{align*}
& \mathcal{O}\bigg(\alpha\cdot \log(p) +\beta\cdot \bigg(\Big(\frac{n^2k}{p}\Big)^{2/3}+ \frac{nk\log(p)}{p}\bigg)
+\gamma\cdot \frac{n^2k}{p}\bigg)
\end{align*}
}\noindent
This gives to the cost recurrence, $T_\mathrm{RT3D}(n,k,p_r^2,n_0)=$
{\mathsmall
\begin{align*}
&\mathcal{O}\bigg(\alpha\cdot \log(p)+\beta\cdot \bigg(\Big(\frac{n^2k}{p_r^2}\Big)^{2/3}+\frac{n^2}{p_r^2}\step{\frac kn}\\
&+\frac{nk\log(p)}{p_r^2}\bigg)\bigg) +\gamma\cdot \frac{n^2k}{p_r^2}\bigg) +2T_\mathrm{RT3D}(\frac{n}{2},k,p_r^2,n_0),
\end{align*}
}\noindent
where we can see that $\frac{nk\log(p)}{p_r^2}=\mathcal{O}((n^2k/p_r^2)^{2/3})$, since the initial partitioning will give $n\geq k$.
It is also easy to see that $\frac{n^2}{p_r^2}\step{\frac kn}=\mathcal{O}((n^2k/p_r^2)^{2/3})$.
With these simplifications, 
{\mathsmall
\begin{align*}
T&_\mathrm{RT3D}(n,k,p_r^2,n_0)
= \mathcal{O}\bigg(\alpha\cdot \log(p)+\beta\cdot \Big(\frac{n^2k}{p_r^2}\Big)^{2/3}\\
&+\gamma\cdot \frac{n^2k}{p_r^2} \bigg) +2T_\mathrm{RT3D}(\frac{n}{2},k,p_r^2,n_0).
\end{align*}
}\noindent
We observe that the bandwidth cost $\mathcal{O}((n^2k/p_r^2)^{2/3})$ decreases by a factor of $2^{1/3}$ at every recursive level, and the computation cost by a factor of $2$.
The base-case cost will be $T_\mathrm{RTBC}(n_0,k,p_r^2)$.
We select $n_0=n^{1/3}\left(\frac{k}{p_r^{2}}\right)^{2/3}$, giving a total cost over all base cases of $\frac{n}{n_0}T_\mathrm{RTBC}(n_0,k,p_r^2)=$
{\mathsmall
\begin{align*}
&\mathcal{O}\bigg(\alpha\cdot \frac{n}{n_0}\log(p)+\beta\cdot \Big(nn_0+\frac{nk\log(p)}{p_r^2}\Big)+\gamma\cdot \frac{nn_0k}{p_r^2} \bigg) \\
&=\mathcal{O}\bigg(\alpha\cdot \Big(\frac{np_r^2}{k}\Big)^{2/3}\log(p)+\beta\cdot \Big(\frac{n^2k}{p_r^2}\Big)^{2/3}+\gamma\cdot \frac{n^{4/3}k^{5/3}}{p_r^{10/3}} \bigg).
\end{align*}
}\noindent
Therefore, the overall cost incurred on each square processor grid is
$T_\mathrm{RT3D}(n,k,p_r^2)=$
{\mathsmall
\begin{align*}
 \mathcal{O}\bigg(\alpha\cdot \Big(\frac{np_r^2}{k}\Big)^{2/3}\log(p)+\beta\cdot \Big(\frac{n^2k}{p_r^2}\Big)^{2/3}
+\gamma\cdot \frac{n^2k}{p_r^2} \bigg).
\end{align*}
}\noindent
When $k\leq n$, we do not have a partitioning step and $p_r^2=p$.
Otherwise, we have $p_r^2=np/k$ obtain the cost $T_\mathrm{RT3D}(n,n,np/k)=$
{\mathsmall
\begin{align*}
 \mathcal{O}\bigg(\alpha\cdot \Big(\frac{np}{k}\Big)^{2/3}\log(p)+\beta\cdot \Big(\frac{n^2k}{p}\Big)^{2/3}
+\gamma\cdot \frac{n^2k}{p} \bigg),
\end{align*}
}\noindent
which is the same as for the case $k\leq n$.
For $n=k$, the 3D costs obtained above are the same as the most efficient algorithms for $n\times n$ LU factorization.
In the subsequent sections, we show that a lower synchronization cost is achievable via selective use of triangular matrix inversion.

\section{Triangular Inversion}
\label{sec:tri}
In this section, we derive the cost of inverting a lower triangular matrix $L$ of size $n \times n$ with $p$ processors.
Since the input matrix is square, the dimensions of the processor grid $\Pi$ should be identical in two dimensions leaving us with $\dim\left(\Pi\right) = p_1\times p_1\times p_2$, where $p=p_1^2p_2$.
We assume the initial matrix to be cyclically distributed on the subgrid $\Pi(\circ,\circ,1)$.

\subsection{Algorithmic Approach}
In \cite{borodin1975computational}, a recursive method for inverting triangular matrices was presented. 
A similar method for full inversion was presented in~\cite{balle1994strassen}.
When applied to a triangular matrix, those methods coincide.
The method uses the triangular structure of the initial matrix to calculate the inverse by subdividing the problem into two recursive matrix inversion calls, which can be executed concurrently and then uses two matrix multiplications to complete the inversion.

Since the subproblems are independent, we want to split the processor grid such that two distinct sets of processors work on either subproblem.
We chose the base case condition to be that the grid is one-dimensional in the dimension of $p_1$ and we do redundant base case calculations in this subgrid.
For this section, we consider $p_2 \geq p_1$, a constraint that we will fulfill anytime the method is called.

{\algsmall
\rule{\columnwidth}{.5pt}
\vspace{-.2in}

\(L^{-1} = \text{RecTriInv}(L,\Pi,p,p_1,p_2)\)

\vspace{-.15in}
\rule{\columnwidth}{.5pt}
\begin{algorithmic}[1]
\Require
\Statex The processor grid $\Pi$ has dimensions $\sqrt{p} \times \sqrt{p}$
\Statex $L$ is a lower triangular $n\times n$ matrix, distributed on $\Pi$ in a cyclic layout, so $\Pi(x,y)$ owns $L[x,y]$ of size $\frac{n}{\sqrt{p}}\times \frac{n}{\sqrt{p}}$ such that $L[x,y](i,j)=L(i\sqrt{p}+x,j\sqrt{p}+y)$. 
\Statex
\If {$p_1=1$}
  \State{$\textbf{AllToAll}\left(L[x,\circ],\Pi(x,\circ)\right) $}
  \State{$L^{-1} = \text{sequential inversion}(L)$}
\Else
  \State{Subdivide $L$ into $n/2\times n/2$ blocks,}
  \State{$L=\begin{bmatrix}L_{11} & 0 \\ L_{21} & L_{22} \end{bmatrix}$}
  \State{Subdivide the processor grid $\Pi=[\Pi_1,\Pi_2]$ such that}
  \markcomment{1}{$\dim(\Pi_1)=\dim(\Pi_2)=(\sqrt{p/2} \times \sqrt{p/2})$}
  \State{Redistribute $\left(L_{11},\Pi \rightarrow \Pi_1\right)$ }
  \State{Redistribute $\left(L_{22},\Pi \rightarrow \Pi_2\right) $ }
  \State{$L^{-1}_{11} = \text{Rec-Tri-Inv}(L_{11},\Pi_1,p,p_1/2^{2/3},p_2/2^{2/3})$}
  \State{$L^{-1}_{22} = \text{Rec-Tri-Inv}(L_{22},\Pi_2,p,p_1/2^{2/3},p_2/2^{2/3})$}
  \State{$L'^{-1}_{21} = -\text{MM}(L^{-1}_{22}, L_{21},\Pi,n,n,p,p_1,p_2)$\label{line:inv_mm1}}
  \State{$L^{-1}_{21} =  \text{MM}(L'^{-1}_{21}, L^{-1}_{11},\Pi,n,,p,p_1,p_2)$\label{line:inv_mm2}}
  \State{Assemble $L^{-1}$ from the $n/2\times n/2$ blocks,}
  \State{$L^{-1}=\begin{bmatrix}L^{-1}_{11} & 0 \\ L^{-1}_{21} & L^{-1}_{22} \end{bmatrix}$\\}
\EndIf
\Statex
\Ensure
\Statex $LL^{-1} = \textbf{1}$ where $L^{-1}$ is distributed the same way as $L$
\end{algorithmic}
\vspace{-.2in}
\rule{\columnwidth}{.5pt}
}

\subsection{Total Cost of Triangular Inversion}
This recursive approach of inverting a matrix has total cost, 
{\mathsmall
\begin{multline*}
  T_\mathrm{RecTriInv}(n,p_1,p_2) = 2T_\mathrm{MM}(n/2,n/2,p_1,p_2) + \\T_\mathrm{RecTriInv}(n/2,p_1/2^{1/3},p_2/2^{1/3}) + T_\mathrm{redistr}(n/2,p_1,p_2), 
\end{multline*}}\noindent
with a base case cost of
{\mathsmall
\[T_\mathrm{RecTriInv}(n_0,1,p_2) = \alpha \cdot 2\log \left(\frac{p_2}{p_1}\right)   + \beta \cdot 2n_0^2 + \gamma\cdot n_0^3.\]
}\noindent
The base case size will be $n_0 = \frac{n}{p_1^{3/2}}$ and therefore neither of the terms is of leading order.
We observe that the bandwidth cost of the matrix multiplication $\mathcal{O}((n^3/p_1^2)^{2/3})$ decreases by a factor of $2^{4/9}$ at every recursive level, and the computation cost by a factor of $2$.
The redistribution process requires moving the matrices from a cyclic processor grid to a smaller cyclic processor grid, with the block each processor owns having a factor of $2^{1/3}$ more rows and columns.
This redistribution is effectively an all-to-all between a larger and a smaller set of processors.
We can get a concrete bound on the cost, by first performing an all-to-all to transition to a blocked layout (each processor owns contiguous blocks of the matrix).
Then we can transition to a blocked layout on the smaller processor grid by scattering each block to at most $4$ processors.
Finally, we can perform an all-to-all on the smaller processor grid to transition from the blocked layout back to a cyclic one.
The overall cost of these steps is $O(\alpha\cdot \log(p) + \beta\cdot n_0^2\log(p)/p)$
This redistribution bandwidth cost is dominated by the cost of the matrix multiplication.

The total cost for the recursive inversion is
{\mathsmall
\begin{align*}
  T_\mathrm{RecTriInv}(n,p_1,p_2) =&  \beta\cdot \frac{2^{1/3}}{2^{1/3}-1}\Big(\frac{n^2}{8p_1^2}+\frac{n^2}{2p_1p_2}\Big)\\
&+\gamma\cdot \frac{2^{1/3}}{2^{1/3}-1}\frac{1}{8}\frac{n^3}{p} + \mathcal{O}\left(\alpha\log^2p\right) .
\end{align*}
}\noindent
In contrast to LU factorization and our recursive TRSM algorithm, the synchronization cost is logarithmic rather than polynomial in p.

\section{Iterative Triangular Solver}
\label{sec:algo}
In this section, we present our main contribution, a 3D TRSM algorithm that uses inversion of diagonal blocks to achieve a lower synchronization cost.
By precomputing the inversions, we replace the latency-dominated small TRSMs with more parallel matrix multiplications.

\subsection{Block-Diagonal Triangular Inversion}
\label{subsec:allinv}
In order to lower the synchronization cost of TRSM, first we invert a set of triangular blocks along the diagonal of the matrix, each with a distinct subset of processors.
We split $\Pi$ into $\frac{n}{n_0}$ subgirds of dimensions $r_1 \times r_1 \times r_2$, where $r_1^2r_2 = p\frac{n_0}{n}$.
To have the proper layout for the inversion, a transition from the original, cyclic layout on a subgrid to the grid as described in Section~\ref{sec:new_trmm} has to happen.
Afterwards, all the inversions can be done in parallel.
To support our inversion, we must have $r_2 > r_1$ and $n_0 \geq \sqrt{r_1^2r_2}$.
The precise choices of $r_1$ and $r_2$ are given in the algorithm and will be discussed in Section~\ref{sec:costanalysis}.

\subsection{Triangular Solve using Partial Inversion}
Initially, we want $L$ to be distributed on the top level of the three dimensional grid $\Pi$ in a cyclic layout such that each processor 
\(\Pi(x,y,1) \text{ owns } L\left(y:p_1: n, x:p_1: n\right).\)
Also, we set the right hand side to be distributed on one level of the grid with a blocked layout with a physical block size of $b \times \frac{k}{p_2}$
such that each processor 
\(\Pi(x,1,z) \text{ owns } B\left(x:p_1:\frac nb, zk/p_2:(z+1)k/p_2\right).\)

\algblock[Name]{Parfor}{EndParFor}
\algblockdefx[NAME]{Parfor}{EndParFor}%
[2][]{\textbf{For} #2 \textbf{do in parallel} }%
{\textbf{end for}}

{\algsmall \rule{\columnwidth}{.5pt} \vspace{-.2in}

\(\tilde{L} = \text{Diagonal-Inverter}(L,\Pi,n,p_1,p_2,n_0)\)

\vspace{-.15in}
\rule{\columnwidth}{.5pt}
\begin{algorithmic}[1]
\Require
\Statex The processor grid $\Pi$ has dimensions $p_1 \times p_1 \times p_2$
\Statex $L$ is a lower triangular $n\times n$ matrix distributed cyclically on $\Pi$ such that processor $\Pi\left(x,y,1)\right) $ owns $L[x,y]$ a lower triangular $\frac{n}{p_1}\times \frac{n}{p_1}$ matrix  such that $L[x,y](i,j)=L(ip_1+x,jp_1+y)$.
\Statex
\State{Define $q=\frac{pn_0}{n}\quad r=\frac{n}{n_0}$}
\State{Define $r_1 = \left(\frac{pn_0}{4n}\right)^{1/3}$}
\State Define $r_2 = \left(\frac{16pn_0}{n}\right)^{1/3}$
\State Define a $p_1\times  p_1\times \sqrt{p_2}\times\sqrt{p_2}$ processor grid $\Pi_{4D}$, such that $\Pi_{4D}(x_1,x_2,y_1,y_2)=\Pi(x_1,x_2,y_1+\sqrt{p_2}y_2)$.
\markcomment{0} and $\Pi_{4D}(x_1,x_2,y_1,y_2)$ owns blocks $L[x_1,x_2,y_1,y_2]$
\State Define a block diagonal matrix $L_D[x_1,x_2,y_1,y_2]$
\markcomment{0}{such that $L_D[x_1,x_2,y_1,y_2][b]$ denotes}
\markcomment{0}{the block $L[x_1,x_2,y_1,y_2](bn_0:(b+1)n_0,bn_0:(b+1)n_0)$}
\State $\textbf{Scatter}\left(L_D[\circ,\circ,y_1,y_2][\circ],\right.$
\markcomment{3}{$\left.\Pi_{4D}(x_1,x_2,1,1),\Pi_{4D}(x_1,x_2,\circ,\circ)\right)$\label{li:scatter}}
\State Define a $\sqrt{p} \times \sqrt{p}$ processor grid $\Pi_{2D}$, such that 
\markcomment{0}{$\Pi_{2D}(x_1 + p_1y_1, x_2+p_2y_2)=\Pi_{4D}(x_1,x_2,y_1,y_2)$.}
\State Define a $\sqrt{q} \times \sqrt{q}\times \sqrt{r}\times \sqrt{r}$ processor grid $\Pi_{4D}^{I}$, such that $\Pi_{4D}^{I}(u_1,u_2,v_1,v_2)=\Pi_{2D}(u_1 + \sqrt{q}v_1, u_2+\sqrt{q}v_2)$ owns blocks $L_D[u_1,u_2,v_1,v_2].$
\State $\textbf{AllToAll}\left(L_D[u_1,u_2,v_1,v_2][\circ],\Pi_{4D}^{I}(u_1,u_2,\circ,\circ)\right)  $\label{li:a2a1}
\Parfor{$i = 0 : \sqrt{\frac{n}{n_0}}-1$}
  \Parfor{$j = 0 : \sqrt{\frac{n}{n_0}}-1$}
    \State{Define $b = \left(i+\sqrt{\frac{n}{n_0}}j\right)$}
    \State{$\tilde{L}_D[\circ,\circ,i,j][b] = \textbf{RecTriInv}\left(L_D[\circ,\circ,i,j][b],\right.$}
    \markcomment{9}{$\left.\Pi_{4D}^{I}\left[\circ,\circ,i,j\right],q,r_1,r_2) \right)$}
  \EndParFor
\EndParFor
\State $\textbf{AllToAll}\left(\tilde{L}_D[u_1,u_2,v_1,v_2][\circ],\Pi_{4D}^{I}(u_1,u_2,\circ,\circ)\right)  $\label{li:a2a2}
\State $\textbf{Gather}\left(\tilde{L}_D[\circ,\circ,y_1,y_2][\circ],\right.$
\markcomment{3}{$\left.\Pi_{4D}(x_1,x_2,\circ,\circ),\Pi_{4D}(x_1,x_2,1,1)\right) $ \label{li:gather}}
\Ensure
\Statex $\tilde{L}_D L_D=\textbf{1} \quad \forall i$ where $L$ and $\tilde{L}$ are partitioned the same way
\end{algorithmic}
\vspace{-.2in}
\rule{\columnwidth}{.5pt}
}

Additionally, each processor has memory of the same size as its part of $B$ allocated for an update-matrix denoted as  \(\overline{B_j},\quad j \in [1,p_1] \), where also each processor $\Pi(x,y,z)$ owns
\(\overline{B_y}\left(x:p_1:\frac nb, zk/p_2:(z+1)k/p_2\right).\)
The algorithm itself consists of two parts: first \bsq{inversion}, we invert all the base-cases on the diagonal in parallel as described in the algorithm above and, second \bsq{solve}, we do the updates and calculate the solution to TRSM.

\section{Cost Analysis of the Iterative TRSM}
\label{sec:costanalysis}
In this section we will derive a performance model for the algorithm presented in Section~\ref{sec:algo}.
The total cost of the algorithm is put together from the cost of its three subroutines:
{\mathsmall
\begin{multline*}
T_\mathrm{It-Inv-TRSM}(n,k,n_0,p_1,p_2) = T_\mathrm{Inv}(n,p_1,p_2) +\\ T_\mathrm{Upd}(n,k,n_0,p_1,p_2) + T_\mathrm{Solve}(n,k,n_0,p_1,p_2).
\end{multline*}}\noindent
Above the cost denoted by inversion is the part of the algorithm that inverts the blocks (Algorithm Diagonal-Inverter).
The solve part is in lines~\ref{line:solveline}-\ref{line:solvereduce}, and the update in lines~\ref{line:updateline}-\ref{line:updatereduce}.

{\algsmall
\rule{\columnwidth}{.5pt}
\vspace{-.2in}

\(X = \text{It-Inv-TRSM}(L,B,\Pi,n,k,p_1,p_2,r_1,r_2)\)

\vspace{-.15in}
\rule{\columnwidth}{.5pt}
\begin{algorithmic}[1]
\Require
\Statex The processor grid $\Pi$ has dimensions $p_1 \times p_1 \times p_2$
\Statex $L$ is a lower triangular $n\times n$ matrix is distributed on $\Pi$ such that $\Pi(x,y,1)$ owns  $L[x,y]$ of size $n/p_1 \times n/p_1$ such that $L[x,y](i,j) = L(ip_1 + x, jp_1 + y)$ 
\Statex $B$ is a dense $n \times k$ matrix is distributed such that $\Pi(x,1,z)$ owns $B[x,z]$of size$n/p_1 \times k/p_2$, such that $B[x,z](i,j) = L(ip_1 + x, zk/p_2 + j)$ 
\Statex Define blocks $S_i = in_0: (i+1)n_0$ and $T_i = in_0 : n$
\Statex
\State{$\tilde{L}= \textbf{Diagonal-Inverter}(L,\Pi,n,p_1,p_2,n_0)$}\label{line:inverseend}
\State{$\textbf{Bcast}\left(B\left[x,z\right](S_0(x),\circ),\Pi(x,1,z),\Pi(x, \circ, z)\right)$}
\For{$i = 0 : \frac{n}{n_0}-1$}
  \State{$\Pi(x,y,z) : X\left[y, z\right](S_i,\circ) = $}
  \markcomment{2}{$\tilde{L}\left[y,x\right](S_i, S_i) \cdot B\left[x,z\right](S_i,\circ) $ \label{line:solveline}}
  \State $X \left[y, z\right](S_i,\circ) = \textbf{Allreduce}\left(X \left[y, z\right](S_i,\circ),\Pi\left(\circ,y,z\right)\right)$ \label{line:solvereduce}
  \State{$\textbf{Bcast}\left(\tilde{L}\left[x, y\right](T_{i+1},S_i),\Pi(x,y,1),\Pi(x,y,\circ)\right)$\label{line:bcastL}}
  \State{$\Pi(x,y,z) : \overline{B_y}\left[x , z\right](T_{i+1},\circ) += $}
  \markcomment{2}{$\tilde{L} \left[x,y\right](T_{i+1},S_i) \cdot X \left[y, z\right](S_i, \circ) $\label{line:updateline}}
  \State $\overline{B_0}\left[x,z\right](S_{i+1},\circ) =$
  \markcomment{2}{$ \textbf{Allreduce}(\overline{B_\circ}\left[x,z\right](S_{i+1},\circ),\Pi\left(x, \circ, z\right))$ \label{line:updatereduce}}
  \State{$\Pi(x,y,z) : B\left[x,z\right](S_{i+1},\circ) = $}
  \markcomment{2}{$B\left[x,z\right](S_{i+1},\circ) - \overline{B_0}\left[x,z\right](S_{i+1},\circ)$\label{line:updatelocal}}
\EndFor
\Ensure
\Statex $B=LX$ where $X$ is distributed the same way as $B$
\end{algorithmic}
\vspace{-.2in}
\rule{\columnwidth}{.5pt}
}

\subsection{Inversion Cost}
We invert the $\frac{n}{n_0}$ submatrices  of size $n_0 \times n_0$ along the diagonal with distinct processor grids.
The size of the processor grids involved  is $r_1 \times r_1 \times r_2$. 
The choices of $r_1$ and $r_2$ are made such that the bandwidth cost of the inversion is minimal.
Additionally we have to account for the cost that arises from communicating the submatrices to the proper subgrids.
This happens in lines \ref{li:scatter}, \ref{li:a2a1}, \ref{li:a2a2}, and \ref{li:gather} of the Algorithm Diagonal-Inverter. The respective costs are summed up the the following table.

{\tablesize
\centering
\begin{tabular}{c | l }
Line~\ref{li:scatter}& $\alpha \cdot \log(p_2) + \beta \cdot \frac{nn_0}{2p_1^2}$ \\\midrule
Line~\ref{li:a2a1}   & $\mathcal{O}\left(\alpha\cdot \log(p)+\beta\cdot \frac{nn_0\log p}{2p}\right) $ \\\midrule
Line~\ref{li:a2a2}   & $\mathcal{O}\left(\alpha\cdot \log(p)+\beta\cdot \frac{nn_0\log p}{2p}\right) $ \\\midrule
Line~\ref{li:gather} &$\alpha \cdot \log(p_2) + \beta \cdot \frac{nn_0}{2p_1^2}$ \\
\end{tabular}

}

These cost are never of leading order compared to the costs that arise form the triangular inversion.
With the derivations done in Section~\ref{sec:tri}, we get the following costs for inversion:

\mypar{Latency Cost}
The total latency cost of inversion is
{\mathsmall  
\[S_\mathrm{Inv}(p) = \mathcal{O}\left(\alpha \log^2p\right). \] 
}\noindent

\mypar{Bandwidth Cost}
In order to minimize the bandwidth cost of triangular inversion, we choose a grid splitting to achieve closest to ideal ratios for the subgrids processor layout $r_1$ and $r_2$.
This ratio is achieved when $r_2 = 4 r_1$. 
The choices for $r_1$ and $r_2$ are therefore,
{\mathsmall  
\[r_1 = \left(\frac{pn_0}{4n}\right)^{1/3} \quad \text{and} \quad 
r_2 = \left(\frac{16pn_0}{n}\right)^{1/3}.\] 
}\noindent
With this grid slicing, we get $ \frac{n}{n_0} $ different sub-grids of dimensions $r_1 \times r_1 \times r_2$.
This setup leads to a cost for inverting $\frac{n}{n_0}$ submatrices of
{\mathsmall
\[
W_\mathrm{Inv}(n_0,r_1,r_2) =\frac{2^{1/3}}{2^{1/3}-1}\left(\frac{n_0^2}{8r_1^2}+\frac{n_0^2}{2r_1r_2}\right).
\]
}\noindent

\mypar{Flop Cost}
The flop cost of the inversion part is
{\mathsmall  
\[F_\mathrm{Inv}(n_0,p_1,p_2) = \frac{1}{8}\frac{nn_0^2}{p_1^2p_2}.\] 
}\noindent

\subsection{Solve Cost}
The complete solve cost can be derived by
{\mathsmall  
\[T_\mathrm{Solve}(n,n_0,k,p,p_1,p_2) = \frac{n}{n_0}T_\mathrm{MM}\left(n_0,k,p,p_1,p_2\right)\] 
}\noindent
\mypar{Latency Cost}
The latency cost of the solve part is
{\mathsmall  
\[S_\mathrm{Solve}(n,n_0,p) = \mathcal{O}\left(\frac{n}{n_0}\log p\right).\] 
}\noindent

\mypar{Bandwidth Cost}
The cost of the solve is one call to triangular matrix multiplication for each base case.
The synchronization cost is again dominated by the $\frac{n}{n_0}$ cases.
The cost for these sums has been presented in Section~\ref{sec:new_trmm}.
Including these, we obtain a total cost of 
{\mathsmall
\begin{multline*}
W_\mathrm{Solve}\left(n,k,n_0,p,p_1,p_2\right)  = \frac{n}{n_0} \cdot W_\mathrm{MM}\left(n_0,k,p,p_1,p_2\right)\\
= \frac{n}{n_0} \cdot \left[\left(\frac{n_0^2}{p_1^2}\right)\step{p_2} + 4 \left(\frac{n_0k}{p_1p_2}\right)\step{p_1}\right].
\end{multline*}}\noindent

\mypar{Flop Cost}
The flop cost of the solve part is
{\mathsmall  
\[F_\mathrm{Solve}(n,k,n_0,p_1,p_2) = \frac{n}{n_0}\left(\frac{n_0^2k}{p_1^2p_2}\right).\] 
}\noindent

\subsection{Update Cost}
The complete solve cost can be derived by 
{\mathsmall  
\[T_\mathrm{Upd}(n,k,n_0,p,p_1,p_2) = \sum_{i=1}^{n/n_0-1} T_\mathrm{MM}\left(n{-}in_0,n_0,k,p,p_1,p_2\right).\] 
}\noindent
\mypar{Latency Cost}
The update latency cost is
{\mathsmall  
\[S_\mathrm{Upd}(n,n_0,p) = \mathcal{O}\left(\frac{n-n_0}{n_0}\log p\right) .\] 
}\noindent

\mypar{Bandwidth Cost}
The cost of doing all the updates as described in the algorithm in Section~\ref{sec:algo} (Lines~\ref{line:solvereduce}-\ref{line:updatelocal}) is the cost of both the allreductions and the broadcast,
{\mathsmall
\begin{multline*}
W_\mathrm{Upd}(n,k,n_0,p_1,p_2) = \\
\sum_{i=1}^{n/n_0-1}\left[W_\mathrm{bcast}\left(\frac{nn_0 - in_0}{p_1^2},p_2\right) + \right.\\
\left. W_\mathrm{allreduction}\left(\frac{n_0k}{p_1p_2}, p_1\right)+W_\mathrm{allreduction}\left(\frac{n_0k}{p_1p_2},p_1\right)\right].
\end{multline*}}\noindent
This yields to a total cost of $W_\mathrm{Upd}(n,k,n_0,p_1,p_2)=$
{\mathsmall
\[
\frac{n - n_0}{n_0} \left[ 4\frac{nn_0-n}{p_1^2}\step{p_2} + 4\frac{n_0k}{p_1p_2} \step{p_1} \right] .
\]}\noindent

\mypar{Flop Cost}
The update flop cost is
{\mathsmall  
\[F_\mathrm{Upd}(n,k,n_0,p_1,p_2) = \frac{n-n_0}{n_0}\left(\frac{knn_0}{p_1^2p_2}\right) .\] 
}\noindent

\subsection{Total Cost}
The total cost of the algorithm is the sum of its three parts and leaves a lot of tuning room as with a choice of $p_1 = 1$, $p_2 = 1$ or $n_0 = n$ one is able to eliminate certain terms. 

\mypar{Latency Cost}
The total latency cost of the algorithm is a sum of the previous parts,
{\mathsmall
\begin{multline*}
S_\mathrm{It-Inv-TRSM}(p_1,p_2,r_1,b) = S_\mathrm{Upd}(n,n_0,p_1,p_2) + \\
S_\mathrm{Solve}(n,n_0,p_1,p_2) + S_\mathrm{Inv}(p_1,p_2,r_1,b)\\
= \mathcal{O}\left(\alpha\left(\frac{n}{n_0}\log p + \log^2p\right) \right).
\end{multline*}}\noindent

\mypar{Bandwidth Cost}
The total bandwidth cost for the TRSM algorithm is, by abbreviating $\nu = \frac{2^{1/3}}{2^{1/3}-1}$,
{\mathsmall
\begin{multline*}
W_\mathrm{It-Inv-TRSM}(n,k,n_0,p_1,p_2,u,v,b)  = \\
W_\mathrm{Upd}(n,k,n_0,p_1,p_2) + W_\mathrm{Solve}\left(n,k,n_0,p_1,p_2\right) \\
+ W_\mathrm{Inv}(n,b,p_1,p_2, u, v)\\
=\frac{n}{n_0} \cdot \left[\left(\frac{n_0^2}{p_1^2}\right)\step{p_2} + 4 \left(\frac{n_0k}{p_1p_2}\right)\step{p_1}\right]  \\
+\frac{n - n_0}{n_0} \left[ 4\frac{nn_0-n}{p_1^2}\step{p_2} + 4\frac{n_0k}{p_1p_2}\step{p_1} \right]+ \nu\left(\frac{n_0^2}{8r_1^2}+\frac{n_0^2}{2r_1r_2}\right) .
\end{multline*}}\noindent

\mypar{Flop Cost}
Lastly, the combined total flop cost is 
{\mathsmall
\begin{multline*}
F_\mathrm{It-Inv-TRSM}(n,k,n_0,p_1,p_2) = F_\mathrm{Upd}(n,n_0,p_1,p_2) \\
+ F_\mathrm{Solve}(n,n_0,p_1,p_2) + F_\mathrm{Inv}(n_0,u,v,p_1,p_2)
= \frac{n^2k}{p_1^2p_2}+\frac{n_0^2n}{p_1^2p_2}.
\end{multline*}}\noindent

\section{Parameter Tuning}
\label{sec:paramtuning}
\begin{figure}[!t]
\centering
\includegraphics[width=0.25\linewidth, angle=-90]{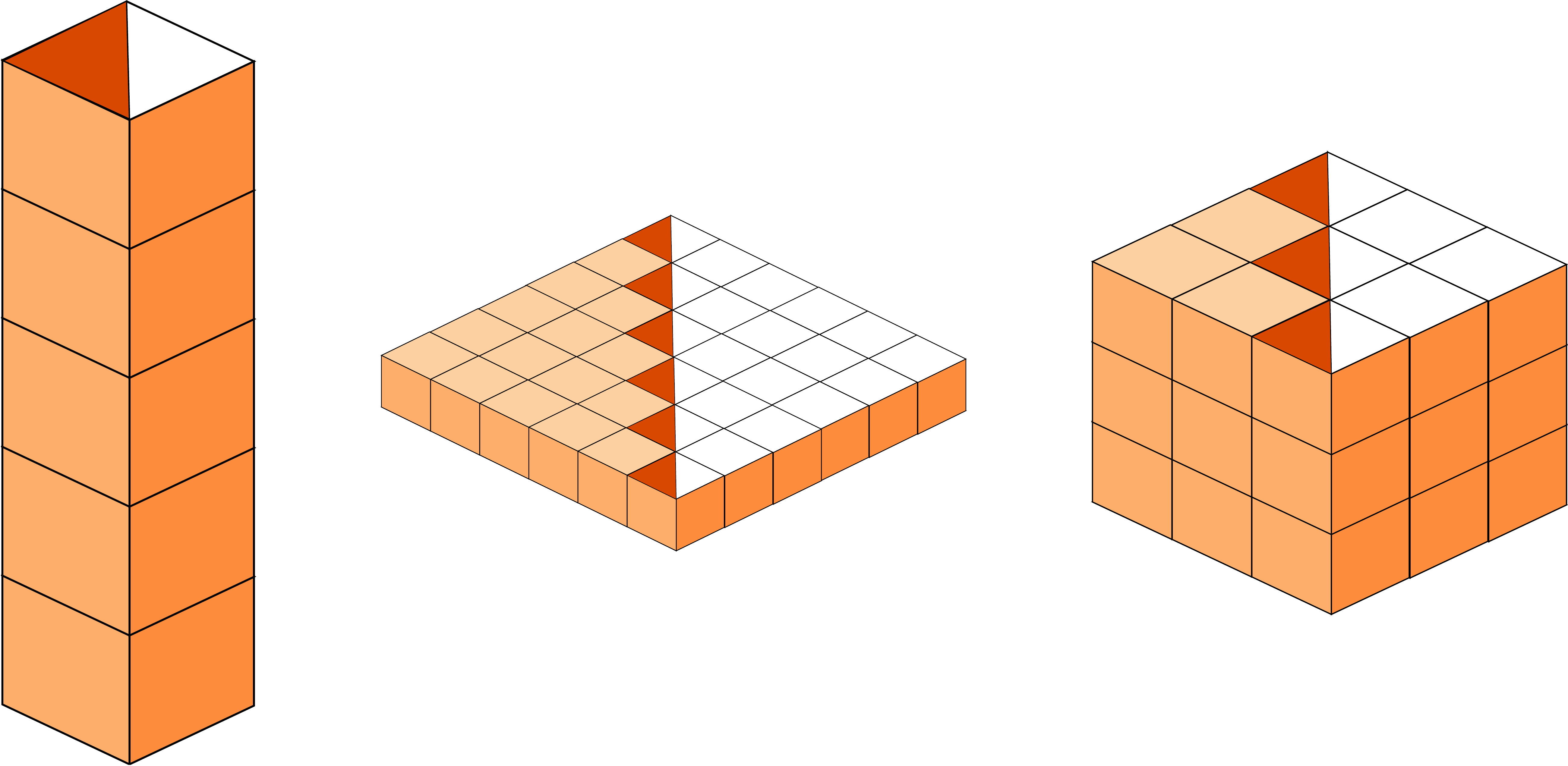}
\caption{One-, two-, and three-dimensional layout dependent on relative matrix sizes. Inverted blocks of the matrix in dark and input- and output of the right hand side on the left and right size of the cuboid.}
\label{fig_inv}
\end{figure}

In this section, we give asymptotically optimal tuning parameters for different relative matrix sizes to optimize performance.
We only focus on asymptotic parameters as there is a trade off between the constant factors on the bandwidth and latency costs.
The exact choice is therefore machine dependent and should be determined experimentally.
The initial grid layout is dependent on the relative matrix sizes of $L$ and $B$ since the update part of the algorithm is one of the dominating terms in any case where there is an update to be made and determines the case where is is infeasible.
The different layouts are shown in Figure~\ref{fig_inv}.

In the case where $n<\frac{4k}{p}$, the processor grid layout is one-dimensional. The optimal parameters are given in the following table.
{\centering
\begin{minipage}{\linewidth}
\centering
{\tablesize
\begin{tabular}{cc || cc || cc}
$p_1 =$ & $1$ &  $r_1 =$ & \multicolumn{1}{c||}{$\mathcal{O}\left(\left(p\right)^{1/3}\right) $}& $n_0 =$ & $n$\\\midrule
$p_2 =$ & $p$ & $r_2 =$ &\multicolumn{1}{c}{$\mathcal{O}\left(\left(p\right)^{1/3}\right) $}\\
\end{tabular}
}
\end{minipage}\hspace{0.1\linewidth}
}
Using this set of parameters will yield to a total cost of
{\mathsmall
\begin{align*}
T_\mathrm{IT1D}(n,k,p) = \mathcal{O}\left(\alpha\cdot \left(\log^2 p + \log p\right) + \beta\cdot n^2 + \gamma\cdot \frac{n^2k}{p}\right). 
\end{align*}
}\noindent
Comparing these costs to the costs of $T_\mathrm{RT1D}$ obtained in Section~\ref{sec:oldcost}, we can see that we are within the asymptotic bounds of the original algorithm in bandwidth and flop cost, but pay an extra factor of $\log p$ latency, since the inversion, if performed on a 3D grid, requires $\log^2p$ steps. But since the inversion is the least significant part of the routine in 1 large dimension, no gain was to be expected in this case.

In the case where $n > 4k\sqrt{p}$, the processor grid layout is two-dimensional. The optimal parameters are given in the following table.
{\centering
\begin{minipage}{\linewidth}
\centering
{\tablesize
\begin{tabular}{cc|| cc ||cc}
$p_1= $ & $\sqrt{p}$ & $p_2=$ & $1$ & $n_0=$ & $\mathcal{O}\left(\left(nk^3p^{1/2}\right)^{1/4}\right)$\\\midrule
$r_1=$ & \multicolumn{3}{c||}{$\mathcal{O}\left(\left(\frac{k}{n}\right)^{1/4}p^{3/8}\right) $}& $r_2=$ &\multicolumn{1}{c}{$\mathcal{O}\left(\left(\frac{k}{n}\right)^{1/4}p^{3/8}\right) $}\\
\end{tabular}
}
\end{minipage}
}
Using this set of parameters will yield to a total cost of
\begin{multline*}
T_\mathrm{IT2D}(n,k,p) = \mathcal{O}\left(\alpha\left(\log^2 p + \left(\frac{n}{k}\right)^{3/4}\frac{1}{p^{1/8}}\log p\right) \right.\\
\left.+ \beta\left(\frac{nk}{\sqrt{p}}\right) + \gamma \left(\frac{n^2k}{\sqrt{p}}\right)\right).
\end{multline*}
Comparing these costs to the cost of $T_\mathrm{RT2D}$ obtained in Section~\ref{sec:oldcost}, we can see that we are asymptotically more efficient in terms of latency by a factor of at least $\frac{p^{1/4}}{\log p}$ as well as in bandwidth by a factor of $\log p$ while having the same flop cost asymptotically. This is a significant gain and especially important as the occurrence of fewer right hand sides $k<n$ is high.

In the case where $\frac{4k}{p} \leq n \leq 4k\sqrt{p}$, the processor grid layout is three-dimensional. The optimal parameters are given in the following table.

{\centering
\begin{minipage}{\linewidth}
{\footnotesize
\begin{tabular}{c|| c ||c}
$p_1= \left(\frac{pn}{4k}\right)^{1/3}$ & $p_2=\left(\frac{\sqrt{p}4k}{n}\right)^{2/3}$ & $n_0=\mathcal{O}\left(\min\left(\sqrt{nk},n\right)\right)$\\\midrule
\multicolumn{3}{c}{\begin{tabular}{c ||c}
\hspace*{-3mm}$r_1=\mathcal{O}\left(\left(\min\left[\frac{p\sqrt{nk}}{n},p\right]\right)^{1/3}\right)$ & \hspace*{-1mm}$r_2=\mathcal{O}\left(\left(\min\left[\frac{p\sqrt{nk}}{n},p \right]\right)^{1/3}\right) $\hspace{-5mm}
\end{tabular}
}

\end{tabular}
}
\end{minipage}\noindent

Using this set of parameters will yield to a total cost of
\begin{multline*}
T_\mathrm{IT3D}(n,k,p) = \mathcal{O}\left(\alpha\left(\log^2 p + \max\left(\sqrt{\frac{n}{k}},1\right) \log p\right) \right.\\
\left.+ \beta\left(\left(\frac{n^2k}{p}\right)^{2/3}\right) + \gamma \left(\frac{n^2k}{\sqrt{p}}\right)\right).
\end{multline*}
}\noindent
Comparing these costs to the cost of $T_\mathrm{RT3D}$ obtained in Section~\ref{sec:oldcost}, we can see that we are asymptotically more efficient in terms of latency by a factor of $\left(\frac{n}{k}\right)^{1/6}p^{2/3}$ while being able to keep bandwidth and flop costs asymptotically constant.

\section{Conclusion}
\label{sec:conclusion}

{\centering
 \footnotesize
\begin{tabular}{l | K{0.38\linewidth} K{0.155\linewidth} K{0.08\linewidth} }
 & S & W & F\\\toprule
 & \multicolumn{3}{c}{1 Large Dimension $\left(n < \frac{4k}{p}\right) $}\\\midrule
standard\hspace*{0.5mm} & \textcolor{green!50!black}{$\log p$} & $n^2$ & $\frac{n^2k}{p}$ \\\midrule
new method & $\log^2 p$ &  $n^2$ &$\frac{n^2k}{p}$ \vspace*{1mm}\\\toprule
 & \multicolumn{3}{c}{2 Large Dimensions $\left(n > 4k\sqrt{p}\right) $}\\\midrule
standard\hspace*{0.5mm} & $\sqrt{p}$ & $\log p \frac{nk}{\sqrt{p}}$ & $\frac{n^2k}{p}$ \\\midrule
new method & \textcolor{green!50!black}{$\log^2 p +\left(\frac{n}{k}\right)^{3/4}\frac{1}{p^{1/8}}\log p$}\vspace*{1mm} & \textcolor{green!50!black}{$\frac{nk}{\sqrt{p}}$} & $\frac{n^2k}{p}$ \\\toprule
 & \multicolumn{3}{c}{3 Large Dimensions $\frac{4k}{p} \leq n \leq 4k\sqrt{p}$}\\\midrule
standard\hspace*{0.5mm} & $\left(\frac{np}{k}\right)^{2/3}\log p$ & $\left(\frac{n^2k}{p}\right)^{2/3}$ & $\frac{n^2k}{p}$  \\\midrule
new method & \textcolor{green!50!black}{$\log^2 p +\sqrt{\frac{n}{k}}\log p$} & $ \left(\frac{n^2k}{p}\right)^{2/3}$ & $\frac{2n^2k}{p}$
\end{tabular}
}

We present a new method for solving triangular systems for multiple right hand sides.
In the above table, we compare to a baseline algorithm adapted from~\cite{gustavson} that achieves costs that are as good or better than the state of the art~\cite{lipshitz_2013,heath1988parallel,Dongarra:1997:SUG:265932}.
Our algorithm achieves better theoretical scalability than these alternatives by up to a factor of $\left(\frac{n}{k}\right)^{1/6}p^{2/3}$. 
For certain matrix dimensions, a decrease of bandwidth cost by a factor of $\log_2 p$ is obtained by use of selective triangular matrix inversion.
By only inverting triangular blocks along the diagonal of the initial matrix, we generalize the usual way of TRSM computation and the full matrix inversion approach.
Fine-tuning the algorithm based on the relative input sizes as well as the number of processors available leads to a significantly more robust algorithm.
The cost analysis of this new method allows us to give recommendations for asymptotically optimal tuning parameters for a wide variety of possible inputs.
The detailed pseudo-code provides a direct path toward a more efficient parallel TRSM implementation.


\bibliographystyle{IEEEtran}
\bibliography{IEEEabrv,paper}
\end{document}